\documentclass[12pt,preprint]{aastex}

\def\ala#1{$^{#1}$} 
\def\BP{Ballesteros-Paredes}

\def\etal{et al.\ }
\def\grados{{$^\circ$}}

\def\HI{HI}
\def\kms{{\rm km~s}$^{-1}$} 
\def\los{LOS}

\def\prommath#1{\langle #1\rangle} 

\def\scf{{\rm SCF}}
\def\scfls{{\rm SCF}$^{l,s}$} 
\def\scfl{{{\rm SCF}$^l$}} 
\def\scfs{{{\rm SCF}$^s$}} 
\def\scfsl{{\rm SCF}$^{l,s}$} 
\def\scf0{{{\rm SCF}$^0$}} 

\def\VS{V\'azquez-Semadeni}

\begin{document}

\title{Velocity Structure of the ISM as Seen by the Spectral 
Correlation Function}

\author{Javier Ballesteros-Paredes$^{1,2,3}$, Enrique
V\'azquez-Semadeni$^2$, and Alyssa A.\ Goodman$^3$}
\affil{$^1$Astrophysics Department, American Museum of Natural History\\
Central Park West at 79th Street. 10024 New York, NY. {e-mail: 
{\tt javierbp@amnh.org}}}
\affil{$^2$Instituto de Astronom\'\i a, Universidad Nacional Aut\'onoma
de M\'exico \\ Apdo. Postal 70-264, 04510 M\'exico D.F., M\'{e}xico.}
\affil{$^3$Harvard-Smithsonian Center for Astrophysics. 60 Garden St.
MS-42. 02138 Cambridge Ma.}

\begin{abstract}
We use the statistical tool known as the ``Spectral 
Correlation
Function" [SCF] to intercompare simulations and observations of the atomic
interstellar medium.  The simulations considered, which mimic three 
distinct sets
of physical conditions, are each calculated for a 300 pc$^3$ box 
centered at the
Galactic plane.   Run ``ISM" is intended to represent a mixture of 
cool and warm
atomic gas, and includes self-gravity and magnetic fields in the calculations.
Run ``ISM-IT" is more representative of molecular clouds, where the gas is
presumed isothermal.  The third run, ``IT" is for purely isothermal gas, with
zero magnetic field, and no self-gravity.  Forcing in the three cases is
accomplished by including simulated effects of stellar heating (for 
ISM), stellar
winds (ISM-IT), or random compressible fluctuations (IT).

For each simulation, H I spectral-line maps are simulated, and it is these maps
which are intercompared, both with each other, and with observations, using the
SCF.  For runs where the separation of velocty features is much 
greater than the
``thermal" width of a line, density-weighted velocity histograms are decent
estimates of H I spectra.  When thermal broadening is large in comparison with
fine-scale turbulent velocity structure, this broadening masks sub-thermal
velocity sub-structure in observed spectra.  So, simulated spectra 
for runs where
thermal broadening is important must be calculated by convolving 
density-weighted
histograms with gaussians whose width represents the thermal broadening.

The H I observations we use here for comparison are of the North Celestial Pole
Loop, a region chosen to minimize line-of-sight confusion on scales $> 100$ pc.
{\it None} of the simulations match the NCP Loop data very well, for 
a variety of
reasons described in the paper.  Most of the reasons for simulation/observation
discrepancy are predictable and understandble, but one is particularly curious:
the most realistic ``simulation" comes from {\it artifically expanding} the
velocity axis of run ISM by a factor of six.  Without rescaling, the high
temperature associated with much of the gas in run ISM causes almost all of the
spectra to appear as virtually identical gaussians whose width is deterimined
solely by the temperature--all velocity structure is smeared out by thermal
broadening.  However, if the velocity axis is expanded $\times 6$, the SCF
distributions of run ISM an the NCP Loop match up fairly well.  This means that
the ratio of thermal to turbulent pressure in run ISM is much too large in the
simulation as it stands, and that either the temperature is much ($\sim 36$
times) lower, and/or that the turbulent energy in the simulation is much too
small.  Run ISM does not include the effects of supernovae, which 
means that the
turbulent energy (and hence velocity scale) is likely to be dramatically
underestimated.

The paper concludes that the SCF is a useful tool for understanding and
fine-tuning simulations of interstellar gas, and in particular that a realistic
simulation of the atomic ISM needs to include the effects of energetic stellar
winds (e.g. supernovae) before the ratio of thermal-to-turbulent pressure will
give spectra representative of the observed interstellar medium in our Galaxy.

\end{abstract}

\section{Introduction}

In the past few decades, the turbulent character of the velocity field
in the interstellar medium has been increasingly recognized (Zuckerman
\&\ Evans 1974; Dickman 1985; Scalo 1987; for recent works, see the
volume by Franco \&\ Carrami\~nana 1999 and the review by
V\'azquez-Semadeni et al. 2000a). Due to the highly fluctuating
character and sensitivity to initial conditions (initial conditions
arbitrarily close to each other separate exponentially in time -- see,
e.g., Lesieur 1990) of the turbulent motions, statistical tools become
important in trying to characterize and understand the nature of
interstellar turbulence. These tools become even more important with
the continuous growth of both observed and theoretical data cubes, as
it becomes necessary to compress such large data sets into a more
manageable parameter space. On the observational side, studies of the
centroid velocity probability distribution function ($v$-PDF) (Miesch
\&\ Scalo 1995; Lis et al. 1996, Miesch, Scalo \&\ Bally 1999) have
compared the shape of the $v$-PDF for various observed regions and
numerical simulations of incompressible or weakly compressible
turbulence; through the density and velocity autocorrelation
functions, Kleiner \&\ Dickman (1984) have tried to infer the
correlation length scale in Taurus Molecular Cloud; using structure
tree statistics, Houlahan \&\ Scalo (1992) tried to discriminate
between hierarchically-nested and random collections of clouds;
through Principal Component Analysis and autocorrelation functions,
Heyer \&\ Schloerb (1997) and Brunt \& Heyer (2001) obtain the
velocity dispersion-size relationship for both observations and
pseudo-simulations, in a manner independent of how clouds are defined;
applying $\Delta-$variance analysis, Stutzki et al.\ (1998) and Bensch
et al.\ (1999) have measured the fractal dimension of the projected
intensity towards the Polaris Flare. Finally, Lazarian \& Pogosyan
(2000) have given a method for recovering the power spectra of both
the three-dimensional density and of the line-of-sight velocity from
the emissivity power spectrum in velocity channels of spectroscopic
data. 

On the numerical side, recent simulations of interstellar turbulence
have allowed statistical studies of the simulated ISM and clouds. In
fact, again due to the sensitivity to initial conditions, modeling of
individual turbulent objects is not feasible, and statistical modeling
is again the appropriate way to proceed. It thus becomes appropriate
to compare numerical simulations and observations using statistical
methods. For example, Falgarone et al. (1994) studied optically thin
line-profiles in 512\ala 3 simulations, concluding that the shapes of
the spectra, their moments, and their spatial variation exhibit
similarities with observational line-profiles of non-star forming
regions; V\'azquez-Semadeni, Ballesteros-Paredes \&\ Rodr\'\i guez
(1997) searched for Larson-type relationships in numerical
simulations, finding in particular that the density-size relationship
is not satisfied, but instead there exists a whole family of
low-column density clouds that may however be missed by observational
surveys limited by integration time (see also Scalo 1990); Padoan \&\
Nordlund (1999) compare sub- and super-Alfv\'enic simulations with
observations, concluding that the first case may be in conflict with
observations; Pichardo et al. (2000) studied the projection of
numerical MHD simulations, finding that the morphology of the channel
maps resembles maps of the line of sight (LOS) of the velocity more
closely than maps of the density field; Mac Low \&\ Ossenkopf (1999)
have recently used the $\Delta$-variance to find that the models with
high Mach numbers (more than M $\sim$ 4) are in better agreement with
observations; Lazarian et al.\ (2001) have shown that the theoretical
predictions of Lazarian \& Pogosyan (2000) are verified for simulated
spectroscopic data from numerical simulations of ISM turbulence.

In particular, Rosolowsky et al.\ (1999 = RGWW) introduced a new
method, called the ``Spectral Correlation function'' (SCF), which
estimates the spatially-averaged\footnote{In this paper, the word
``spatial'' refers to the plane of the sky, unless otherwise stated.}
correlation of spectra inside a box of size $s$ in spectroscopic data
cubes. These authors used the SCF to measure the correlation between
spectra in neighboring positions, as an indicator of the degree of
``smoothness'' of both observational and numerical data cubes. With
this method, they were able to distinguish between, for example,
simulations with and without self-gravity, by comparing the variation
of the SCF upon randomization of the positions of the spectra on the
plane of the sky (POS). 

In the present work, we extend the work of RGWW, further exploring the
capabilities of the SCF to discriminate among numerical simulations of
various turbulent regimes (isothermal with large- and small-scale
forcing, and ISM-like), and to quantitatively compare them to
spectroscopic observational data of diffuse HI gas. In order to do
this, we construct spectral maps from the numerical data cubes
assuming optically thin lines, with and without thermal broadening of
lines. For non-isothermal gas, the velocity structure of the colder
gas may be ``hidden'' by the thermal broadening of the warmer gas. 

The plan of the paper is as follows: in \S \ref{datos} we describe
both the observed  HI data (Hartmann \&\ Burton 1997 ) and the
simulated data cubes used here, and explain the construction of the
spectral maps for the simulations. In \S \ref{spectral} we briefly
review the SCF. In \S \ref{resultados} we apply the SCF to the data
sets, discussing the information that can be extracted from this
procedure towards distinguishing between the various simulations and
quantitatively estimating their similarity with observational data. In
\S \ref{discussion} we discuss the nature of the spatial structure of
the SCF, and discuss how the SCF can be used to guide theoreticians to
match their simulations with observational data. Finally, in \S
\ref{conclusions} we draw the main conclusions. 

\section{Data}\label{datos}

\subsection{Numerical Simulations}\label{datos_simulaciones}

We consider three numerical simulations in three dimensions. The first
one, called simply ISM, represents the ISM in a box of 300 pc on a
side, at a resolution of 100 points per dimension, centered on the
Galactic midplane at the solar Galactocentric distance. Full
equations, details and a complete description of the model can be
found in Passot et al. (1995 =PVP) and Pichardo et al. (2000). Here we
just mention that the 
code solves the full self-gravitating MHD equations, including that of
the internal energy conservation  with additional model terms
representing the background plus local stellar heating and the
radiative cooling. The local heating mimics stellar ionization
heating, and is turned on (``an O star is formed'') when the local density
reaches a threshold value of 10 times the mean density,
producing local bubbles of warm gas that expand and feed the global
turbulence. ``Stars'' remain on for 2 Myr. No supernova-like energy
injection is included because of numerical limitations. The cooling is
parameterized as a piecewise
power-law function, as fitted to the cooling functions of Dalgarno \&
McCray (1972) and of
 Raymond, Cox \&\ Smith (1976) by Rosen, Bregman \&\ Norman (1993)
and Rosen \&\ Bregman (1995) (see PVP  and Pichardo et al. 2001). The
``isothermal'' Jeans length is twice the integration box size,
although the presence of cooling gives an effective Jeans length of
roughly 2/3 the box size (\BP, \VS, \&\ Scalo 1999). The mean density is 
3.3 cm${-3}$ and the velocity, temperature and magnetic field units are, 
respectively, 6.4 km s${-1}$, 3000 K and 5 $\mu$G. 

Two limitations of the ISM simulation should be mentioned. First of all,
the lack of SN energy input implies that the simulations are somewhat
less energetic than the actual ISM. This fact will be quantified in
\S\ref{fine_tune}. On the other hand, the heating and 
cooling functions used do not produce any thermally unstable temperature
ranges, although we do not consider this to be a serious problem, as
recent discussions by V\'azquez-Semadeni et 
al. (2000b) and Gazol et al. (2001) suggest that the structural
properties of the ISM are more a consequence of the dynamical 
processes than of the thermal ones. On the other hand, a true
shortcoming is the absence of hot ($T \sim 10^6$ K) gas, although recent
work by various groups (e.g., Gazol-Pati\~no 
\& Passot 1999; de Avillez 2000) suggests that the filling factor of the 
this gas at the midplane is not larger than $\sim 20$\%. In this
case, the error commited by not including it is probably minor, as the main 
effect of this omission will be that in actual observations a fraction
of the line of sight will contain ``holes'' in the cold/warm gas
distribution, filled with hot ionized gas,
which will be absent in our simulation data. Given the 
low resolution of our simulations, this may actually constitute an
advantage, as better sampling of the gas contributing to the HI line
spectra along each LOS will be accomplished.

The second simulation, called ISM-IT, is similar to run ISM, except
that an isothermal regime is considered, with no cooling and heating
terms, the isothermal Jeans length is 1.1 times the box size, and a
lower density threshold is used for initiating star formation, so that
the star formation rate is similar to that of run ISM in spite of the
weaker gravity. In this run, which may be thought to represent gas
within a molecular cloud at $T\sim$~10~K, the stellar energy input
proceeds via winds (locally divergent source terms in the momentum
equation -- see \VS\ et al.\ 1996; Avila-Reese \& \VS\ 2001), and the
internal energy evolution equation is bypassed altogether, but
self-gravity and the magnetic field are still included. Finally, the
third run, also at a resolution of $100^3$ and called IT, is a purely
hydrodynamic isothermal run without self-gravity and with random
compressible forcing applied at scales of 1/4 the simulation box
size. As a summary, we present in Table~\ref{Table:sim} the main
features of each one of the runs. 

\begin{deluxetable}{l|l|l|c|l}
\tabletypesize{\scriptsize} 
\rotate
\tablecaption{ \label{Table:sim}}
\tablewidth{0pt}
\tablehead{	\colhead{Run name and description}
		&\colhead{Physical Conditions}
		&\colhead{Synthetic line spectra generated}
		&\colhead{No. of Vel Channels}
		&\colhead{Comments}
}
\startdata
	 \rm{ISM}
	&\rm{300pc$^3$ box}
	&\rm{NTB: Density-weighted}
	&\rm{16 and 64}
	&\rm{Unrealistic, spiky spectra. These}
\\ 
	 \rm{Represents cool and warm }
	&\rm{Galactic midplane}
	&\rm{  velocity histograms}
	&\rm{}
	&\rm{structures are unobservable}
\\ \cline{3-5}
	 \rm{neutral HI gas}
	&\rm{Conditions at $R_\odot$}
	&\rm{TB: Each cell broadened by}
	&\rm{16}
	&\rm{Spectra too smoth and uniform}

\\ 
	
	&\rm{Self gravitaty: on}
	&\rm{  temperature at that position}
	&
	&\rm{across the map}
\\ \cline{3-5}
	
	&\rm{Magnetic Fields: on}
	&\rm{Velocity adjusted ($\times$ 6)}
	&\rm{16}
	&\rm{More realistic spectra and SCF.}
\\ 
	
	&\rm{Lj $\sim$ 200 pc}
	&
	&
	&\rm{Shows that ISM run is less}
\\ 
	
	&\rm{Forcing: Stellar heating}
	&
	&
	&\rm{turbulent than HI data}
\\ 
	
	&\rm{Density threshold for SF:}
	&
	&
	&
\\ 
	
	&\rm{$\rho_{\rm th} = 8$~cm$^{-3}$}
	&
	&
	&
\\ \hline
	 \rm{ISM-IT}
	&\rm{Same as ISM but:}
	&\rm{NTB}
	&\rm{16}
	&\rm{Non-realistic spectra.}
\\ 
	 \rm{Isothermal, but with ISM}
	&\rm{Isothermal}
	&
	&
	&\rm{Still ``wormy'' appareance}
\\ 
	 \rm{ingredents. Represents}
	&\rm{Lj(isothermal) ~ 1.1 box size}
	&
	&
	&\rm{of the SCF maps}
\\ 
	 \rm{molecular clouds.}
	&\rm{Forcing: stellar ``winds''}
	&
	&
	&\rm{}
\\ 
	
	&\rm{Lower density threshold}
	&
	&
	&
\\ 
	
	&\rm{for SF: $\rho_{\rm th} = 4$~cm$^{-3}$}
	&
	&
	&
\\ \hline
	 \rm{IT}
	&\rm{Pure HD, no self-gravity}
	&\rm{NTB}
	&\rm{16}
	&\rm{Non-realistic spectra}
\\ 
	 \rm{Isothermal}
	&\rm{Isothermal }
	&
	&
	&\rm{Roundish ``worms'' in the SCF}
\\ 
	
	&\rm{Forcing: Random, compressible}
	&
	&
	&\rm{caused by the forcing.}
\\ \hline

\enddata
\end{deluxetable}


All three runs are started with Gaussian fluctuations with random
phases, uncorrelated among all variables. Note that
for run IT, the forcing is active since 
the beginning of the simulation, while for runs ISM and ISM-IT, the
forcing is turned on only after the conditions for star
formation mentioned above are met. 


\subsubsection{Observing the Simulations}

In order to produce ``spectra'' from the simulations,  we generate two
types of line profiles from the simulations. In one, we
simply calculate  density-weighted $z$-velocity histograms 
(position-position-velocity  cubes, or PPV). This is equivalent to 
assuming that the emission is optically thin, but also that the thermal
broadening is negligible, so that what we see is directly the
emission from each pixel at its physical velocity along the LOS. Such
spectra are produced by considering all points $(x_0,y_0,z)$ over
the $z$-coordinate, and then creating a velocity axis by
adding the contribution from each $z$-position to a velocity bin
corresponding to its LOS-velocity ($u_z$). We refer to these
as not-thermally-broadened (NTB) spectra.  These synthetic observations, 
although unsuitable for direct comparison with real observational data
because the latter can not escape some level of thermal broadening, 
allow us to study the effects of velocity structure exclusively.

For the second type of spectra, we calculate
the radiative transfer taking the density field in the
simulation to correspond to the HI density, and assuming
local thermodynamic equilibrium, optically thin lines, and Gaussian
thermal broadening. In this case, every pixel ``emits'' at all
velocities, but its emission is weighted by a Gaussian profile centered 
at the fluid velocity of the pixel, and of width determined by the
temperature at the pixel.
We will refer to this kind of spectra as thermally-broadened (TB)
spectra. In both cases, ``channel maps'' are then
produced as intensity images over the remaining two spatial coordinates
($x$ and $y$) at a given $u_z$ interval.

In Fig.~\ref{ejemplos_lineas} we show the corresponding TB and NTB 
spectra for nine pixels around position $(x=3,y=3)$ of run ISM. The
dotted lines
represent the TB spectra, while the bold solid lines represent the 
NTB spectra at low (16 velocity channels) resolution, and the light
solid lines represent the NTB spectra at high velocity resolution (64
channels). It is seen that 
the NTB spectra are much more irregular and spiky, because there are
only 100 events in each \los, which are then redistributed
in either 64 (at high velocity-resolution) or 16 (at low velocity-resolution)
possible bins. Note also that the
spectra change substantially between the low- and high-resolution cases. In
contrast, the TB spectra are much smoother,
and they do not change so much with position on the POS, implying that
thermal broadening ``hides'' the real velocity structure of the
simulations. We will return to this point in \S\ref{therm_broad}.

It is worth pointing out that previous works constructing velocity
histograms out of 3D numerical simulations have frequently added 
up the contribution from several neighbouring \los s. For example, 
Falgarone \etal (1994), using $512^3$ simulations, considered data 
from a square box of 32 $\times$ 32 neighbouring \los s, thus 
including 524288 events for each individual spectrum. However, 
at the risk of being velocity under-sampled, in the
present work we use only one \los\ per spectrum, in order to avoid
further deterioration of the already low on-the-sky spatial resolution
imposed by the machine available to us (a Cray Y-MP 4/64).

\subsection{Observational Data}\label{observaciones}

As the source of observational data, we use the 21~cm HI data from the
Atlas of Galactic Neutral
Hydrogen (Hartmann \&\ Burton 1997) obtained with the 25~m Leiden/Dwingeloo
telescope. Details of the observations and analysis may be found in the
atlas. Here we just mention that the velocity resolution is 1.03 \kms,
and that we use the velocity range from $-$20 to $+$20 \kms\ (the
original data was taken from $-$450 to 400 \kms.) 

We have chosen the North Celestial Pole Loop as the comparison region.
This is an extended HI region located at $120 \leq l \leq 160$,
$10\leq b \leq 50$ (see \S~\ref{caso_npl}.) There are several reasons
for this choice, mostly related to the fact that the ISM simulation
represents a well-defined region of space, while HI observations suffer
from confusion along the LOS. First, it is a region that lies off the Galactic
plane, thus avoiding excessive contribution to the emission from very distant,
physically unrelated regions. Secondly, the presence of a well-defined
physical structure (the NCP Loop)
reassures us that the emmision is mostly due to a well-confined region
is space. Finally, there are molecular clouds and
stars that allow to have an estimation of the distance, at least to
some of the molecular clouds in the region (MBM 30, $d=110\pm 10$~pc),
as well as an upper limit ($d=346$~pc) to other members (MBM 29; see
Penprase 1993). This implies that, if the emission lies at a distance
of $\sim$~230~pc (the mean between these two extremes), the dimensions
of the analyzed region has a size $\sim$160~pc,
comparable to the physical size of the ISM simulation box. 

Finally, it is worth noting that the resolution of the \HI\ data
changes with galactic latitude: it is  0.5\grados\ for $b \leq$
35\grados; 0.6\grados\ for 35\grados $\leq b <$ 45\grados; and
0.7\grados\ for 45\grados $\leq b  <$ 55\grados. This has the effect
of slightly overestimating the SCF at high latitudes, since we assume
a uniform grid spacing of 0.5\grados.

\section{The Spectral Correlation Function}\label{spectral}

A detailed discussion of the development and advantages of the SCF is
given by RGWW. For completeness, here we
repeat the necessary definitions, give a physical interpretation of
the SCF, and then a discussion of the strategies for its
application, and the effects of noise. 

\subsection{Definitions and Physical Interpretation}\label{physical}

Given two arbitrary spectra with antenna temperatures $T_1(v)$ and
$T_0(v)$, the SCF is defined by RGWW as:

\begin{equation}
{\rm SCF} \equiv 1-\sqrt{{\int [s T_1(v-l) - T_0(v) ]^2 dv\over
s^2\int T_1^2(v-l)dv + \int T^2_0(v)dv}}
\label{scf} 
\end{equation} 
where $l$ and $s$ are free parameters chosen as to minimize the integral
$\Upsilon=\int [s T_1(v-l) - T_0(v) ]^2 dv$. Thus, for two identical
spectra with infinite signal to noise (s/n) ratio, SCF$ = 1$, while
for two completely uncorrelated spectra, SCF = 0.

Given the parameters $l$ and $s$, RGWW define four {\it modes}
of the SCF:
a) \scf0 $\equiv$ SCF$(s=1,l=0)$, which compares the two spectra in
their raw form. Values of \scf0 close to unity imply that the spectra
resemble each other in shape, amplitude and velocity offset.
b) \scfl $\equiv$ SCF$(s=1,l)$, with $l$ chosen as to minimize
$\Upsilon$. This form eliminates differences between
the spectra arising from pure velocity offsets. Values of \scfl close
to unity imply that the spectra resemble each other in shape and
amplitude, regardless of velocity offset.
c) \scfs $\equiv$ SCF$(s,l=0)$, with $s$ chosen as to minimize
$\Upsilon$. This form eliminates differences between
the spectra arising from pure intensity scalings. Values of \scfs close
to unity imply that the spectra resemble each other in shape and
velocity offset, regardless of absolute amplitude.
d) \scfsl $\equiv$ SCF$(s,l)$, with both $s$ and $l$ chosen as to 
minimize $\Upsilon$. This is then the general definition
of the SCF (eq.\  [\ref{scf}]), and eliminates differences arising
from both velocity offset and amplitude. In this case, values close to 
unity imply that the two spectra resemble each other in shape only.

\subsection{Effects of Instrumental Noise}\label{ruido}

As has been shown by RGWW (see also Padoan, Rosolowsky and Goodman 
2001 = PRG), the SCF can be strongly affected by noise and their 
effects has to be evaluated. PRG propose to ``clean'' 
the effects of the noise by dividing the actual value of the SCF by an
``optimal value'' of the SCF, i.e., by the maximum value that the SCF 
would have if the only difference between both spectra were the random 
noise. The optimal value is given by (see PGR)

\begin{equation}
{\rm SCF}(Q) = 1-{1\over Q},
\label{renormalization}
\end{equation}
where $Q$ is the ``spectrum quality'', defined as

\begin{equation}
Q_i = {1\over \sigma}\sqrt{{\Sigma_v T^2_i \delta v \over N_{\rm ch} \delta v}}
= {1\over \sigma} \sqrt{\prommath{T^2_i}},
\label{Qu}
\end{equation}
where in turn $\sigma$ is the mean rms noise, $N_{\rm ch}$ is the number 
of velocity channels, $\delta v$ is their width, and $\prommath{T^2_i}$ is the 
mean value of the emission in the velocity interval.
As can be seen, for high values of the signal-to-noise ratio, 
the parameter $Q$ goes to infinity, and the SCF is renormalized by a 
number close to unity.

PGR show that adding noise to both the observed (molecular, with
signal-to-noise values of the order of a few) and the synthetic (without
intrinsic noise) spectra until the signal-to-noise ratio is constant
throughout the data, and renormalizing the SCF as indicated in
eq.\ (\ref{renormalization}), gives values of the SCF essentially independent
of the noise level.

In the present work we deal with data that are almost not affected by 
noise: on one hand, the numerical simulations 
are intrinsically noiseless; on the other, for the HI data 
(Hartmann \&\ Burton 1997, see \S\ref{datos}), the characteristic 
value of $\sigma$ is 
0.07~K and its characteristic antena temperature value is of several K,
suggesting high values of $Q$. 
Indeed, we have verified that for the observed data set (\HI) used
here, the mean value of $Q$ is 98.8 (see \S\ref{caso_npl}). Therefore,
in this paper we do not correct for the effects of noise.

\subsection{Velocity Window}\label{window}

In order to compute the four kinds of the SCF, in particular, the modes
\scfls\  and \scfl, which imply  shifting one spectrum with respect to the
other, RGWW use velocity 
channels within 3 FWHMs of the velocity centroid, with the FWHM and the 
velocity centroid computed by fitting a Gaussian to the line 
profiles. In turn, PRG use a 10~$\Delta_2 v$ velocity window, where 
$\Delta_2 v$ is the standard deviation of the velocity of a spectrum created 
by averaging over the whole map. However, in our case of negligible
noise, we simply extend
the original data sets (observational and synthetic) by adding zeros on 
both sides of the spectral lines to produce a spectral window three
times wider than the original.

\subsection{Maps and Histograms of the SCF}\label{scf_caja}

The SCF, like the regular autocorrelation function, is in general a
function of spatial separation, and this aspect of it is dicussed by
PRG. However, RGWW has used it in a
``local'' mode, which we follow here. In this form, we construct maps of
the average value of the SCF in  
boxes of 3 pixels per side in a given spectral map. That
is, to each point on the $(x,y)$ plane (the POS of the
simulations) we
assign the average value of the SCF evaluated using that point and
each of its eight nearest neighbors. Note that with this
approach, the SCF maps are useful for exhibiting small-scale spatial 
variations in the line spectra. In addition, 
in order to quantify the similarity between spectral maps, RGWW
also considered the distribution (histogram) of the SCF in a
given map and its moments, and we follow this approach here. 

In practice, the nearest-neighbor variant of the SCF used here acts like
an angle-average of the spatial (i.e., in the 
POS) derivative of the spectral profiles. Sites of strong profile 
variation appear in the
SCF maps as low values (dark), and soft variations appear as
large values (bright). That is, the SCF stresses small-scale changes (on 
the POS) in the {\it line spectra}. Moreover, the fact that the
SCF considers the whole velocity structure in the line profiles at
each position, implies that maps of the SCF give more information than
the channel maps alone, because they make {\it velocity} variations
(as well as density ones) evident to the eye, which might escape
examination of pairs of channel maps. The SCF histogram moments then
quantify such velocity {\it and} intensity differences.

\section{Results and Interpretation}\label{resultados}

\subsection{Understanding the SCF}\label{tests section}

In order to ``calibrate'' the SCF, and to determine its response upon
changes in the spectral resolution and thermal broadening, we have
simulated four different ``observations'' of run ISM, labeled 16-TB, 16-NTB,
64-TB and 64-NTB, where the number indicates the number of velocity
channels, and the letters TB/NTB indicate whether or not thermal
broadening was included. We also have produced another set (TB and NTB) of
``observations'' of this run, but with the velocity artifitially multiplied by
a factor of 6, in order for the line spectra and the SCF histograms to
better match those for the NCP Loop. This 
will be discussed in \S\ref{fine_tune}.  Velocity channel maps, maps of the 
SCF, and SCF histograms are respectively shown in figs.\ \ref{ISM_vel_chan}, 
\ref{ISM maps scf} and \ref{ISM_hist}. Each case (16/64 velocity 
channels, and TB/NTB) is indicated on the top of each figure.

\subsubsection{Effects of the velocity resolution} \label{vel_res}

Comparing the NTB channel maps (Figs.~\ref{ISM_vel_chan}b and d), we
note that at higher velocity resolution the small-scale structure is
more prominent, a fact already noted by Pichardo et al.\ (2000), and
interpreted by Henney, \VS\ \& Pichardo (2001) in terms of the presence 
of ``caustics'' or ``cusps'' (loci of high intensity due to velocity
crowding) in the position-position-velocity (PPV) data cube. We will return
to these caustics in \S \ref{worms_caustics}. However, this effect is
almost non-existent in the more realistic TB case. 

Additionally, it can be seen that the NTB map at low velocity (or
``spectral'') resolution 
(fig.~\ref{ISM maps scf}b) exhibits a clear ``wormy'' or ``stringy'' 
structure, probably related to the caustics mentioned above (\S
\ref{fine_tune}. However, the wormy structure is almost absent from the
high-resolution SCF maps (fig~\ref{ISM maps scf}d).
This is an artifact of the low {\it spatial} resolution of our
simulations (100 grid points per dimension), which causes 
any given point in the simulated POS to have
very few points along the LOS contributing to the
intensity in any one velocity channel (less than two on average at a velocity
resolution of 64 channels), in turn causing the line profiles to be
highly irregular (cf.\ fig.\ \ref{ejemplos_lineas}). Thus, the 
little signal available is diluted over many velocity channels, and only
the most promintent features are preserved, with a very 
``blurred'' quality. This shows that the effective velocity resolution is 
in effect limited by the spatial resolution. 
The scattered character of the NTB line profiles at high velocity resolution
is also evident in their corresponding SCF histograms (fig.\ 
\ref{ISM_hist}d), which are clearly much wider and have lower mean values than
the low-resolution ones (fig.\ \ref{ISM_hist}b), indicating that the
profiles vary strongly
among neighboring LOS. 

\subsubsection{Effects of thermal broadening} \label{therm_broad}

We now turn to the effects of thermal broadening on velocity channel
and SCF maps, and on the SCF histograms. For the discussion, we
consider only the low-velocity resolution observations, due to the poor
statistics present in the high-resolution ones. 

It is first important to note that run ISM contains large
regions occupied by warm ($T \sim 1.5\times 10^3$ K 
gas, whose thermal velocity 
dispersion dominates the line profiles at any given position on the
POS, as seen in fig.\ \ref{ejemplos_lineas}. This is because, even
though the low-temperature regions (clouds) contain supersonic motions, these
motions are subsonic compared to the sound speed of the warm
medium. Moreover, the warm gas is barely transonic with respect to
itself. Thus, {\it the thermal broadening due to the warm gas along the LOS 
masks the bulk velocity structure} seen in the profiles, which are
nearly perfectly Gaussian in the TB case, in contrast with the
extremely irregular profiles seen in the NTB case. This effect is
expected to occur in actual observations of multi-temperature atomic
gas, so that the velocity structure of the cold HI gas may only be seen in
absorption (Gibson et al.\ 2000; Heiles 2001).

This effect is also seen in the channel maps. The structure in the NTB channel 
maps, regardless of resolution, can be described as having the appearance
of superposed veils, with sharp, bright edges delimiting fainter
regions. In contrast, the TB channel maps are much smoother, losing
the veil-like appearance, and becoming more similar to maps of the
total integrated intensity. Thus, only the gross features such as
large-scale velocity gradients remain noticeable. 
Again, this effect is also expected to occur in actual observational data
of warm gas.

In the SCF maps, the ``wormy'' nature of the NTB cases, mentioned in \S
\ref{vel_res}, is seen to lose its small-scale part in the TB
maps. That is, the ``worms'' in the NTB case appear to ``bundle up''
into larger-scale filaments, in a sort of self-similar or hierarchical 
manner. However, in the TB SCF maps, only the bundles appear to be
preserved, while their smaller constitutive filaments are not seen. As 
will be discussed in \S \ref{discussion}, the small-scale filaments,
if originated by the caustics in the PPV cube, would be an artifact of the
velocity segregation of the data upon the process of a spectroscopic
observation. Thus, thermal broadening has the fortunate effect of
counteracting this spurious small-scale generation by the velocity
separation effected by the observation process. 

Another important feature is that the SCF maps for the TB
observations have a much smaller dynamic range than the NTB ones,
indicating a much lower variability of the SCF in the former case
although, on the other hand, the same overall structures are seen in
both types of maps. The SCF histograms clearly reflect the above effects. 
While the histograms of the NTB observations have large variances (fig.\
\ref{ISM_hist}b) and moderate mean values, those for the TB (fig.\
\ref{ISM_hist}a) cases are much narrower and with means very close to unity,
the maximum possible value of the correlation. The small
variances and large means of the TB cases again indicate the ``erasure'' of the
velocity structure by the thermal broadening, which causes all spectra 
to be much more similar to each other than in the NTB cases. This
furthermore indicates that the structure seen in the TB SCF maps
should be due mostly to variations in intensity rather than in
velocity structure. This is indeed confirmed by the fact that, in the
TB cases, the
histograms of \scfs, which can compensate for intensity variations but 
not for velocity shifts, are much tighter (have smaller variances)
than those of \scfl, which compensates in the opposite
manner. Instead, in the NTB cases, the histograms of \scfs\ have
comparable (in the low velocity resolution case) or even larger (in
the high-resolution case) variances than the histograms of \scfl,
indicating that in this case the variability in the maps is due mainly to the
velocity structure.

\subsection{Comparison between numerical simulations}\label{other
simulations}

We now compare these results with two other numerical simulations. As 
mentioned in \S \ref{datos_simulaciones}, we consider an isothermal
simulation including the magnetic field and self-gravity (called
ISM-IT) on one hand, and a simple isothermal simulation (called IT)
on the other.  Given the strong masking effect of thermal 
broadening discussed in the 
previous section, in this section we discuss only NTB
cases, so that the velocity structure is fully accesible to the
SCF. However, note that this cannot be done in the case of real
observational data.

Figures \ref{IT_vel_chan}a, and b show the velocity channels for
cases ISM-IT and IT, respectively. The veil-like structure
described for run ISM is also present in runs IT and ISM-IT, although
in the latter the veils are more roundish, somewhat resembling soap
bubbles, or peanut shells. This is most likely a reflection of the
intermediate-scale forcing applied to run IT (in scales of about 
1/4 of the box), in contrast with the
small-scale forcing acting on runs ISM and ISM-IT, due to the stellar
sources. The presence of the
magnetic field in runs ISM and ISM-IT is probably also a factor
contributing to the more elongated structures present
there\footnote{Note that in the ISM simulations, $\beta$, the magnetic to 
thermal pressure ratio, is less meaningful than in isothermal simulations,
since both the magnetic and thermal pressure are not constant and, in fact, 
$\beta$ is a local quantity, spanning a factor of at least 3 orders of
magnitude.}.

Another evident distinction between the runs is that very-large-scale
structures (runing diagonally from one side of the box to the other)
of roughly half of the box size), which are present in run 
ISM (see e.g., velocity channel at $\sim 4$~\kms\ in fig~\ref{ISM_vel_chan}b), 
are absent from cases IT and ISM-IT. This is because of the smaller
effective Jeans length in run ISM (\S \ref{datos_simulaciones}). Thus,
such large-scale 
structures in run ISM have a gravitational origin, but have no
counterpart in runs ISM-IT and IT.

The SCF maps of runs ISM-IT and IT (figs.\ \ref{IT_maps_scf}a and b)
continue to show the ``wormy'' 
appearance described for run ISM, although the ``worms'' form rounder
bundles in run IT, due to the also rounder shapes of the
structures seen in its channel maps. 
The SCF histograms for all three runs (figs.\ \ref{ISM maps scf}b and
\ref{IT_maps_scf}a and b) have comparable means and
variances. However, the histograms for
runs in which star formation is active (runs ISM and ISM-IT) seem to
have more extended low-correlation tails, and therefore more negative
values of the skewness. This appears to be a consequence of the greater
abundance of shells and shocks in cases with active star-formation,
and thus suggests that {\it the histograms of the SCF are capable of
``sensing'' the presence of these structures,
representing them by large negative skewness values}. Unfortunately, the 
correspondence is not unique: the phenomena that can cause a
large low-SCF tails are not limited to shocks and shells. Any
sharp changes in the gas structure, either in
density or in LOS-velocity, or even in the PPV cubes (e.g., the
caustics), can cause them. Thus, the method cannot be 
used to unambiguously identify shocks and shells. Instead, the
existence of low-SCF tails in the histograms should be taken only as
suggestive of their possible presence, which should be then verified
by other means. 

\subsection{SCF for the North Polar Loop}\label{caso_npl}

We now apply the SCF to the HI maps of the North Celestial Pole (NCP)
Loop, a region possibly formed by a nearby supernova explosion (see, e.g.,
Pound \&\ Goodman 1997). In fig.~\ref{npl}a we show the channel maps for
this region, which contains an arc-shaped structure appearing at positive
velocities, and a clear large-scale velocity gradient: at large
negative velocities most of the emission is concentrated at high
longitude and low latitude, while at large positive velocities most of 
the emission comes from the arc and is concentrated in the upper half of
the region.

As we mentioned before (\S\ref{observaciones}), the reason for
choosing this region is that it is a high-latitude region, avoiding
the contribution to the emission from physically disconnected regions
far away in the line of sight, and that the determination of the
distance to some molecular clouds allows us to assume that the size of
the region is about 160~pc, comparable to the size of the simulations
(300~pc). 

Figure \ref{npl}b then shows the SCF maps for this region. Several
points are apparent. First, the ``wormy'' structure seen
in all the simulations is also seen in the NCP Loop, particularly in 
the arc region, indicating the presence of shocks or caustics in the
field (\S\S 
\ref{vel_res} and \ref{discussion})\footnote{The ``worms'' in the 
observations' SCF maps at first sight appear larger than those in 
the simulations' SCF
maps. However, this is only due to the larger pixel size of the NCP Loop
maps. The filaments are in all cases only a few pixels wide.}. In
this case, it is natural to assume that it is primarily shocks that are
causing the ``worms'', the NCP Loop being an expanding shell. Second,
we see that the SCF maps 
only contain structure around the arc, and not near the bottom, where
rather featureless emission is seen in the channel maps, indicating that
{\it the local form of the SCF used in this paper eliminates roughly
uniform regions, and 
highlights regions of strong variability}. Another consequence of
this behavior is that {\it the local form of the SCF eliminates
large-scale velocity 
gradients, as indicated by the fact that no sign of the gradient seen in
the channel map set is apparent in the SCF maps}. Although this is in a sense
obvious, it makes the local SCF maps a powerful tool for studying the
small-scale intensity and velocity structure. In particular, the SCF should
make a good shell-identification tool, in the sense that we can
quantify, using the skewness of the histograms, the relative
importance of sharp edges (shells) of gas.

Also worth noting is the fact that the map of \scfl\ contains
significantly lower values of the correlation compared to the map of
\scfs, indicating that most of the variability is due to the intensity
rather than to the velocity structure. This is also seen in the SCF
histogram for \scfl\ (fig.\ \ref{npl}c), which shows a more extended
low-correlation tail, and consequently a larger negative skewness than 
the \scfs\ histogram. However, these features may instead be an 
indication that thermal broadening is not negligible for this region,
thus partially ``blinding'' the SCF to the velocity structure. If this is so,
then the thermal broadening should be of 3-5~\kms, since the total width of
the HI line profiles for this region is about 8-10~\kms, such that the total
width comes from the turbulent structure, but the thermal broadening does not
allow to see in detail such turbulent velocity structure. Indeed,
it is noteworthy that the histograms of \scfl\ for the simulations have
larger negative skewnesses than those of \scfs\ in the thermally
broadened cases, but the opposite is true for the NTB cases, supporting
this view. This is 
also probably the reason why the featureless emission at the bottom of 
the channel maps does not contain ``worms'' associated to the
small-scale velocity structure in the SCF maps.

\section{Discussion}\label{discussion}

\subsection{``Worms'' as shocks or caustics in PPV space} 
\label{worms_caustics} 

The local form of the SCF as used in this paper is by definition
sensitive only to small-scale 
features (either in intensity or in velocity).
Specifically, structural features in the SCF maps correspond to
{\it steep gradients in the plane of the sky} 
of the line profiles' shape and/or amplitude. In
particular, shocks and the caustics discussed by Henney et al.\ (2001) 
are especially well suited for detection by the local SCF.

For clarity, a brief summary of the results of Henney et al.\ (2001)
is in order here. Essentially, the process of performing a
spectroscopic observation amounts to performing a mapping from the
real, three-dimensional physical space onto position-position-velocity 
(PPV) space (see also Lazarian \& Pogosyan 2000; Pichardo et al.\
2000). As is well known, ``caustics'' form in PPV 
space at places where the mapping is not single-valued. i.e., where
finite-length line-of-sight segments contribute to infinitesimal
velocity intervals, causing a diverging intensity at those sites in
the PPV cube (a phenomenon known as ``velocity crowding''). With 
finite-width velocity channels, the intensity does not diverge, but
still reaches large values at the caustics. The latter are surfaces in 
the PPV cube, and in general intersect the velocity channels
(constant-velocity planes in the PPV cube) at finite angles along
lines which show up as elongated, ``worm''-like features in the channel
maps. This structure is spurious in the sense that it does not 
correspond to any actual density features in the original physical space. In
the case of sharply-defined velocity channels, the spurious
structure reaches arbitrarily small scales as the velocity resolution
is increased. Upon thermal
broadening, the minimum scale present is explicitly calculated by
Henney et al.\ (2001).

This spurious small-scale structure may however be
picked out by the local SCF, showing up as the ``worms'' we have
mentioned repeatedly. However, if
the thermal broadening is very strong (i.e., the velocities are strongly
subsonic), the caustics become thicker and less intense in PPV space, so 
that the ``worms'' become wider and fainter. Thus, as already mentioned in \S
\ref{therm_broad}, thermal broadening present in real spectroscopic 
observations has the fortunate effect of
counteracting the spurious small-scale generated by the synthetic observation
process, although at the cost of blurring the velocity information.

The other possible type of small-scale feature that is naturally
picked out by the local SCF is shocked shells. Contrary to the caustics,
shocks are real features in 3D space, and, since they imply
supersonic velocity differences occurring over very small scales, it is 
possible that they ``survive'' thermal broadening. We conclude that,
when thermal broadening is not negligible, ``worms'' in SCF maps are
most likely to correspond to real velocity features --
shocks. However, for extremely supersonic motions where thermal
broadening is negligible, it is possible that the SCF may not be able
to distinguish between them and the spurious caustics. Further work is
necessary to quantify and distinguish caustics and shocks through the
SCF.

\subsection{``Diagnosing'' the simulations through the SCF}\label{fine_tune}

The SCF is intended to provide a quantitative means of comparing
numerical simulations and observations, which should then allow
workers to adjust their simulations to make them match the observational 
data as closely as possible. In this section we show an example of how this 
can be done.

Comparing both the NTB and the TB 16-channel SCF histograms of the ISM
simulation (figs.\ \ref{ISM_hist}a and b) with those of the NCP Loop (fig.\
\ref{npl}c), we note that neither set of the simulation histograms is a
very good match of those of the NCP Loop, and instead seem to bracket
the latter. The NTB histograms are too broad and have smaller means than the 
NCP Loop ones, while the opposite is true of the TB histograms. In
particular, the narrowness of the TB histograms suggests that the thermal width
dominates the line spectra of the simulation, causing them to be
very similar, all being very close to Gaussians. Indeed, as
seen in fig.~\ref{ejemplos_lineas}, the thermally broadened profiles
have lost essentially all the velocity information and are strongly
dominated by the thermal component. Instead, the 16-channel NTB
profiles are very spiky and do not look like realistic profiles.

As mentioned in \S \ref{therm_broad}, we attribute this problem to the
fact that,
even though the cold gas contains supersonic motions, these motions are
not supersonic compared to the thermal speed of the warm gas, and thus
the velocity information of the cold gas is swamped in the thermal width 
of the warm gas. This is a real effect, and we expect it to affect the
NCP Loop data. However, it moreover occurs that the latter 
include the contribution from gas moving at significantly
larger velocities, most likely due to large shell expansion velocities (see 
Pound \&\ Goodman 1997). Instead, the simulation only contains motions 
in the interval $-6$ km s$^{-1} \lesssim v \lesssim 6$ km s$^{-1}$,
as it includes only HII region-like expanding shells,
but not supernova remnant-like ones, which would have much larger
velocities. Thus, in the HI data there exists quite a larger
velocity range, which ``escapes'' the thermal smoothing.
This suggests that, in order to make the simulated
line profiles more realistic, we should ``expand'' the LOS-velocity axis by
some factor, until the ``right'' ratio of non-thermal velocity range to
thermal width is obtained. The expansion factor can be determined by
matching the SCF histograms of the simulation with thermal broadening
and velocity expansion to those of the HI data. We find that expanding
the velocity axis by a factor of 6 gives the best match, as shown in
figs.\ \ref{sim_fix}a and b, which respectively present the SCF histograms
for the thermally-broadened, velocity-expanded simulation data and
some selected line profiles. Note also that, since the line profiles of
fig.\ \ref{sim_fix}b appear significantly more realistic, this figure
also shows the
extent to which the velocity structure is hidden by the thermal
broadening, by comparing the solid lines (thermally broadenend spectra)
to the dotted lines (actual velocity histograms along the LOS).

Note that the matching can still be improved, as other differences
between the simulation and the NCP Loop data still remain at the next level
of refinement: first, although we have matched the mean and standard
deviation of the SCF histograms, the skewnesses and kurtoses of the
simulation data are considerably smaller than those of the NCP Loop
data. This is a reflection that in the NCP Loop channel SCF maps there exist 
regions of very low values of the SCF corresponding to the arc,
which is a strong and large feature, while no such prominent structures
are present in the simulation data. Second, the histograms of \scfl\
have larger (negative) skewnesses than the histograms of \scfs, while
the opposite is true for the simulation histograms. As mentioned in \S
\ref{caso_npl}, we have interpreted this as a consequence that in the
NCP Loop SCF maps the variability has a larger contribution from intensity
variations, while in the simulation data there seems to be a larger
contribution from velocity variation along an LOS. Thus, it seems that
better matches could be obtained by comparing the simulation data to a
region of sky without such a prominent feature, although for brevity we do not
pursue this further here, as we believe that the potential of the SCF to 
guide the simulations to match the observational data has been
sufficiently exemplified.

Certainly, the velocity expansion we have done here is not a
self-consistent fix
to the simulation, as the correct thing to do would be to re-scale the
whole simulation to Galactic scales, and to correctly take into account
the disk rotation and stratification, as well as more powerful sources
of energy (supernovae) capable of accelerating the gas to higher velocities. 
The conclusion is that our simulations represent a much less turbulent medium
than the actual ISM,
as implied by their smaller velocity dispersion compared to the velocity
dispersion of the H~I gas. This exercise has shown the potential of the
SCF to pick out differences between 
the HI and simulation PPV data, and to point towards the physical
ingredients that are lacking in the simulations for them to be better
models of a given type of observation.

\section{Summary and Conclusions} \label{conclusions}

In this paper we have presented further discussion on the Spectral
Correlation Function (SCF), in addition to that given by 
RGWW and PRG, and then applied it to
data from numerical simulations of the ISM and from HI data (Hartmann
\& Burton 1997) of the North Celestial Pole (NCP) Loop. The application
to the numerical data, for which 
all the information on the distribution of the physical variables is
known, allows us to understand the response of the various SCF
measurements used here on the velocity structure in the simulations. This
understanding was then used to interpret the results of applying the 
SCF to the NCP Loop data, and to suggest modifications to the parameters of
the ISM simulation to better match the observational data.

In this paper, as in RGWW, the SCF has been used in a ``local''
form, in which the SCF between one position in a map and its nearest
neighbors is calculated and assigned as an ``intensity'' to that
position, thus producing {\it maps} of the SCF. Thus, structural
features in the SCF maps correspond to {\it sharp gradients} 
in the line profiles' shape and/or amplitude on the plane of the sky.
Also, histograms of the
SCF values in the maps have been studied. This procedure has been
applied to all four ``modes'' of the SCF, \scf0, \scfs, \scfl\ and
\scfsl. The behavior of the SCF with separation, similarly to the more 
familiar case of regular correlation functions in hydrodynamics (see,
e.g., Lesieur 1990), is discussed by PRG.

We first reviewed basic definitions concerning the SCF, in particular
the four ``modes'', which allow the SCF to focus on spectral
similarities in either total intensity, velocity structure, none, or both.
We then considered the effect of noise, and showed that in
the case of HI (and certainly in the case of numerical simulations) it
is not an issue, allowing us to neglect its effects in the subsequent
discussion.

In the absence of thermal broadening, the SCF maps of all simulations
exhibited a strikingly ``wormy'' structure, with an interesting kind
of self-similarity in which the smallest-scale ``worms'' bundle up to
form larger-scale ones. The ``worms'' may be associated 
to caustics in the position-position-velocity (PPV) data cube (loci of high
intensity due to velocity crowding), discussed by Henney et
al.\ (2001). Such caustics are not a real feature of the physical
fields, but are generated by the sepctroscopic observation process,
and they are
certainly present in the PPV cube, strongly influencing the
structure seen in the channel maps. Their presence causes the SCF
histograms to broaden, and to have lower mean values and larger
negative skewnesses. 

Application of the SCF to a multi-temperature, self-gravitating MHD
simulation of the ISM at intermediate scales (3-300 pc), showed that
thermal broadening may dominate the spectral lines when the warm
gas is primarily subsonic, even if cold, locally supersonic regions
exist along the line of sight (assuming optically thin lines), as
these regions are still sub- or trans-sonic with respect to the warmer
gas. This caused the small-scale structure (which also
typically consists of smaller velocity amplitudes) to disappear from the 
thermally-broadened (TB) SCF maps compared to the non-thermally
broadened (NTB) ones, and the TB SCF histograms to be much narrower
and with higher mean values than their NTB counterparts. This implies
that in the case of the multi-temperature HI data much of the
small-scale velocity structure associated with the colder gas may be
unobservable in emission, and is possibly only accessible through
self-absorption features in the line profiles.



We then applied the SCF to two isothermal runs, one including stellar
energy injection, self-gravity and the magnetic field (run ISM-IT),
which may be representative of molecular gas, and 
one without all these ingredients and forced at large scales (run IT).
We did not consider thermal broadening for these runs because
nearly isothermal (molecular) gas in the ISM is generally very
supersonic, at least above scales $\sim 0.1$ pc.
The channel maps for these runs were seen to be markedly different
from those for 
run ISM, lacking the largest-scale structures formed by large-scale
gravitational instability in the latter run (run ISM-IT  
has a Jeans length larger than the simulation size, and run IT has no
gravity at
all). However, SCF maps of the isothermal runs were not found to be so
different, because the ``local'' SCF used in this paper focuses on
small-scale structure.

The SCF histograms of simulations ISM and ISM-IT, which have
active star-forming regimes and thus contain stronger, more numerous
shocks, tend to have more extended tails at low values of the SCF, and
thus large negative skewnesses. This appears to be due to the strong
spectral and intensity variations caused by the shocks, suggesting that
large negative skewnesses in the SCF
histograms is indicative of the presence of shocks and/or shells in the
physical fields. Unfortunately, other structures in the gas, or even artifacts 
in the PPV cube such as the caustics, can also produce such low-SCF tails in
the histograms, and thus the indication cannot be completely unambiguous.

Then, SCF maps and histograms of the NCP Loop region were
analyzed. The ``wormy'' structure is seen in the maps for this case also,
indicating that the caustics are a real concern. Moreover, the
histograms also exhibit large negative skewnesses, due to the prominence
of the shell-like structure in this region, in agreement with the
suggestion from the simulation data.

We then proceeded to show how the SCF can be used to
``fine tune'' the simulations to better match the observations. Indeed,
the means and standard deviations of the NTB and TB simulation
histograms are seen to bracket the corresponding values for the NCP Loop
histograms, and in fact the TB line profiles are seen to be completely
thermally dominated, with virtually no velocity information left. 
Attributing this effect to the fact that the ISM simulation contains a
rather limited velocity contrast, we were able to obtain a much better
match to the NCP Loop SCF histograms, and much more realistic line profiles, 
by ``expanding'' the velocity range by a factor of 6, indicating that
the simulation requires stronger sources of energy to generate larger
velocity dispersion. In other words, we found that the simulations
presented here are less energetic than the gas in the NCP region.

We conclude that the local mode of the SCF used here has the ability
to quantify the presence of specific velocity and structural features in 
observational data such as shocks/shells, as well as to allow a
quantitative comparison between observational and simulation data.

\acknowledgements We are pleased to acknowledge C. Heiles for
enlightening and enjoyable discussions.  This work
was supported in part by NASA Astrophysical Theory Program grant
no. NAG5-10103 and  CONACYT grants 88046-EUA to J. B.-P., and 
27752-E to E.V.-S.

{}

\begin{figure} 
\caption{Nine line profiles for run ISM centered at position (3,3) on
the simulation's ``plane of the sky'' (the [$x,y$] plane). The {\it
solid-bold lines} represents the not thermally broadened (NTB) line
spectra at a resolution of 16 channels. The {\it solid-thin} lines
represent the NTB spectra with 64 velocity channels, and the {\it dotted
lines} represent the thermally broadened (TB) spectra at 16 velocity
channels.  Note that in the latter all the spectral features are hidden
by the thermal broadening, and that the high-resolution spectra are
poorly sampled, as evidenced by their spiky nature.
\label{ejemplos_lineas}}
\end{figure}



\begin{figure}
\end{figure}

\begin{figure}
\end{figure}

\begin{figure}
\end{figure}

\begin{figure} 
\caption{Velocity channel maps for the four different ``observations''
of run ISM. a) 16 velocity channels, TB; b) 16 velocity channels, NTB;
c) 64 velocity channels, TB; d) 64 velocity channels, NTB. NTB maps show
veil-like structures with sharp edges. TB maps are much smoother.
\label{ISM_vel_chan}}
\end{figure}


\begin{figure} 
\end{figure}

\begin{figure} 
\end{figure}

\begin{figure} 
\end{figure}

\begin{figure} 
\caption{SCF maps for the four different ``observations'' of run ISM
shown in Fig.~\ref{ISM_vel_chan}. Several point are noteworthy: i) TB
SCF maps ({\it a and c}) highlight places where shocks and
shells occur (dark regions); ii) NTB SCF maps ({\it b and d}) exhibit a
``wormy'' structure probably associated to caustics in the PPV data
cube; iii) high resolution SCF maps from TB data ({\it c}) do not show
important differences with respect to the low-resolution TB case; iv) high
resolution SCF maps from NTB data ({\it d}) exhibit a subtantially less
``wormy'' structure than low resolution NTB data.
\label{ISM maps scf}}
\end{figure}



\begin{figure} 
\end{figure}

\begin{figure} 
\end{figure}

\begin{figure} 
\end{figure}

\begin{figure} 
\caption{Histograms of the SCF maps shown in fig.~\ref{ISM maps scf}. 
Since the correlation between neighbouring spectra is larger in the TB
case than in the NTB one, the histograms for the former have larger mean
values and smaller widths 
and skewnesses than the latter. Also, a high velocity resolution, in the 
absence of strong thermal broadening, reduces the correlation, giving
wider histograms with lower means.
\label{ISM_hist}}
\end{figure}

\clearpage


\begin{figure} 
\end{figure}

\begin{figure}
\caption{NTB channel maps for: a) the isothermal, self-gravitating run
with point-like energy (run ISM-ITa);
b) the isothermal, purely hidrodynamic run with random forcing at 
intermediate scales (1/4 of the box) (run IT). Note that the 
veil-like structure seen in the NTB channel maps of run ISM is also
present in these simulations, but not so the
very-large-scale structures (of the size of the box) seen in that run.}
\label{IT_vel_chan}
\end{figure}


\begin{figure} 
\end{figure}

\begin{figure} 
\caption{SCF maps for figs. \ref{IT_vel_chan}. The wormy structure seen
in run ISM is also present, but bundling into more roundish forms,
following the structures that appear in the channel maps (Fig
\ref{IT_vel_chan}). 
\label{IT_maps_scf}}
\end{figure}


\begin{figure} 
\end{figure}

\begin{figure} 
\caption{SCF histograms for figs. \ref{IT_vel_chan}. Although both runs 
have similar mean and variance, the run with star formation ({\it a}) exhibits 
more extended low-correlation tails, suggesting that the SCF is capable to 
pick up multiple shocks and shells.
\label{IT_hist}}
\end{figure}



\begin{figure} 
\end{figure}

\begin{figure} 
\vskip 1cm
\end{figure}

\begin{figure} 
\caption{{\it a)} Channel maps; {\it b)} SCF maps, and {\it c)} SCF
histograms for the 
North Celestial Pole (NCP) Loop. As in the simulations, a ``wormy'' 
structure is seen in the maps, and low-SCF sites denote
the places where shell-like structure is seen.
\label{npl}} 
\end{figure}


\begin{figure}
\end{figure}

\begin{figure}
\caption{(a) Line profiles of run ISM at the same positions as in
Fig.~\ref{ejemplos_lineas} but with the velocity axis multiplied 
by a factor of 6. The thermally-broadened line profiles
now appear more realistic than either the NTB or TB profiles of fig.\
\ref{ejemplos_lineas}. (b) The SCF histograms are much closer in mean and
standard deviation to those of the NCP Loop data than either the NTB or the
TB histograms shown in figs.\ \ref{ISM_hist}a and b.}
\label{sim_fix}
\end{figure}


\begin{thebibliography}{}

\bibitem[Avila]{Avila} Avila-Reese, V.\ \& \VS, E., 2001, ApJ, 553, 645

\bibitem[Avillez]{Avillez} de Avillez, M. A. 2000, MNRAS 315, 479

\bibitem[Ballesteros-Paredes, V{\'a}zquez-Semadeni, \& 
Scalo(1999)]{1999ApJ...515..286B} Ballesteros-Paredes, J., 
V{\'a}zquez-Semadeni, E., \& Scalo, J.\ 1999, \apj, 515, 286 

\bibitem[Dickman(1985)]{1985prpl.conf..150D} Dickman, R.\ L.\ 1985, 
in Protostars and planets II.
Tucson, AZ, University of Arizona Press, 1985, p. 150-174.

\bibitem[Falgarone et al.(1994)]{1994ApJ...436..728F} Falgarone, E., Lis, 
D.\ C., Phillips, T.\ G., Pouquet, A., Porter, D.\ H., \& Woodward, P.\ R.\ 
1994, \apj, 436, 728 

\bibitem[Franco \& Carraminana(1999)]{1999intu.conf.....F} Franco, J.\ \& 
Carraminana, A.\ 1999. Interstellar Turbulence, Proceedings of the 2nd 
Guillermo Haro Conference. Cambridge University Press.

\bibitem[Gazol \& Passot 99]{GP99} Gazol- Pati\~no, A. \& Passot,
T. 1999, ApJ 518, 748

\bibitem[Gazol et al 2001]{GVSS01} Gazol, A., \VS, E.,
S\'anchez-Salcedo, F. J. \& Scalo, J. 2000, ApJ 557, L121

\bibitem[Gibson]{Gibson} Gibson, S.J., Taylor, A. R., Higgs, L. A., Dewdney,
P.E. 2000, ApJ, 540, 85

\bibitem[Hartmann \& Burton(1997)]{1997agnh.book.....H} Hartmann, D.\ \& 
Burton, W.\ B.\ 1997. Atlas of galactic neutral hydrogen. 
Cambridge; New York: Cambridge University Press.


\bibitem[Heiles (2001)]{Heiles01} Heiles, C. 2001, in Galactic
Structure, Stars, and the Interstellar Medium, eds. M.D. Bicay \& C.E.\
Woodward (San Francisco: ASP), in press (astro-ph/0010047)

\bibitem[Henney]{Henney00} Henney, W \VS\ \& Pichardo, B. 2001. In preparation

\bibitem[Heyer \& Schloerb(1997)]{1997ApJ...475..173H} Heyer, M.\ H.\ \& 
Schloerb, F.\ P.\ 1997, \apj, 475, 173 

\bibitem[Houlahan \& Scalo(1992)]{1992ApJ...393..172H} Houlahan, P.\ \& 
Scalo, J.\ 1992, \apj, 393, 172 

\bibitem[Kleiner \& Dickman(1984)]{1984ApJ...286..255K} Kleiner, S.\ C.\ \& 
Dickman, R.\ L.\ 1984, \apj, 286, 255 

\bibitem[Lazarian \& Pogosyan(2000)]{2000ApJ...537..720L} Lazarian, A.\ \& 
Pogosyan, D.\ 2000, \apj, 537, 720 

\bibitem[Lazarian]{lazetal01} Lazarian, A., Pogosyan, D.,
V\'azquez-Semadeni, E., \& Pichardo, B. 2001. Lazarian, A., Pogosyan,
D., V{\'  a}zquez-Semadeni, E., \& Pichardo, B.~;.\ 2001, \apj, 555,
130

\bibitem[Lesieur 1990]{Lesieur90}Lesieur, M. 1990. Turbulence in
Fluids. Kluwer Academic Publishers.

\bibitem[Lis, Pety, Phillips, \& Falgarone(1996)]{1996ApJ...463..623L} Lis, 
D.\ C., Pety, J., Phillips, T.\ G., \& Falgarone, E.\ 1996, \apj, 463, 623 

\bibitem[Mac Low \& Ossenkopf(2000)]{2000A&A...353..339M} Mac Low, M.\ -.\ 
\& Ossenkopf, V.\ 2000, \aap, 353, 339 

\bibitem[Miesch \& Scalo(1995)]{1995ApJ...450L..27M} Miesch, M.\ S.\ \& 
Scalo, J.\ M.\ 1995, \apjl, 450, L27 

\bibitem[Miesch, Scalo, \& Bally(1999)]{1999ApJ...524..895M} Miesch, M.\ 
S., Scalo, J., \& Bally, J.\ 1999, \apj, 524, 895 

\bibitem[Padoan \& Nordlund(1999)]{1999ApJ...526..279P} Padoan, P.\ \& 
Nordlund, {\AA}ke 1999, \apj, 526, 279 

\bibitem[Padoan, Rosolowsky, \& Goodman(2001)]{2001ApJ...547..862P} Padoan, 
P., Rosolowsky, E.\ W., \& Goodman, A.\ A.\ 2001, \apj, 547, 862 

\bibitem[Pichardo et al.(2000)]{2000ApJ...532..353P} Pichardo, B.\ ;., 
V{\'a}zquez-Semadeni, E., Gazol, A., Passot, T., \& Ballesteros-Paredes, 
J.\ 2000, \apj, 532, 353 


\bibitem[Pound \& Goodman(1997)]{1997ApJ...482..334P} Pound, M.\ W.\ \& 
Goodman, A.\ A.\ 1997, \apj, 482, 334 

\bibitem[RGWW]{RGWW} Rosolowsky, Goodman, A.  A., Wilner, D. 1999, ApJ 
524, 887

\bibitem[Scalo87]{Scalo87} Scalo, J. 1987, in Interstellar processes; 
Proceedings of the Symposium, Grand Teton National Park, WY, July 1-7, 1986
Dordrecht, D. Reidel Publishing Co., 1987, p. 349-392.

\bibitem[Scalo(1990)]{1990ppfs.work..151S} Scalo, J.\ 1990, ASSL Vol.\ 162: 
Physical Processes in Fragmentation and Star Formation, Dordrecht, 
Netherlands, Kluwer Academic Publishers, 1990, p. 151-176

\bibitem[Stutzki et al.(1998)]{1998A&A...336..697S} Stutzki, J., Bensch, 
F., Heithausen, A., Ossenkopf, V., \& Zielinsky, M.\ 1998, \aap, 336, 697 

\bibitem[Vazquez-Semadeni, Ballesteros-Paredes, \& 
Rodriguez(1997)]{1997ApJ...474..292V} Vazquez-Semadeni, E., 
Ballesteros-Paredes, J., \& Rodriguez, L.\ F.\ 1997, \apj, 474, 292 

\bibitem[Vazquez-Semadeni et al.(2000a)]{2000prpl.conf....3V} 
V\'azquez-Semadeni, E., Ostriker, E.\ C., Passot, T., Gammie, C.\ F., \& 
Stone, J.\ M.\ 2000a, Protostars and Planets IV (Book - Tucson: University 
of Arizona Press; eds Mannings, V., Boss, A.P., Russell, S.\ S.), p.\ 3, 3 

\bibitem[Vazquez-Semadeni et al.(2000b)]{VGS00} V\'azquez-Semadeni, E.,
Gazol, A. \& Scalo, J. 2000, ApJ 540, 271

\bibitem[Vazquez-Semadeni, Passot, \& Pouquet(1996)]{1996ApJ...473..881V} 
Vazquez-Semadeni, E., Passot, T., \& Pouquet, A.\ 1996, \apj, 473, 881 


\bibitem[Zuckerman \& Evans(1974)]{1974ApJ...192L.149Z} Zuckerman, B.\ \& 
Evans, N.\ J.\ 1974, \apjl, 192, L149 


\end{thebibliography}
\end{document}